\documentclass[pre,twocolumn,amsmath,amssymb,floatfix,,superscriptaddress]{revtex4-1}

\usepackage{graphicx}
\usepackage{dcolumn}
\usepackage{bm}

\usepackage{lipsum}

\usepackage[FIGTOPCAP,tight,raggedright,nooneline]{subfigure}
\usepackage{float}
\graphicspath{{images/}{plots/}}

\newcommand{\avg}[1]{\ensuremath{\left< #1 \right>}}

\begin{document}

    \title{Large deviations of a random walk model with emerging territories}
    \author{Hendrik Schawe}
    \email{hendrik.schawe@cyu.fr}
    \affiliation{Laboratoire de Physique Th\'{e}orique et Mod\'{e}lisation, UMR-8089 CNRS,
        CY Cergy Paris Universit\'{e}, 95000 Cergy, France}
    \author{Alexander K. Hartmann}
    \email{a.hartmann@uol.de}
    \affiliation{Institut f\"ur Physik, Universit\"at Oldenburg, 26111 Oldenburg, Germany}
    \date{\today}

    \begin{abstract}
        We study an agent-based model of animals marking their territory and evading
        adversarial territory in one dimension, with respect to the distribution of the size of
        the resulting territories. In particular, we use sophisticated sampling methods
        to determine it over a large part of territory sizes, including atypically small
        and large configurations, which occur with probability of less than $10^{-30}$.
        We find hints for the validity of a large deviation principle, the shape of
        the rate function for the right tail of the distribution and insight into
        the structure of atypical realizations.
    \end{abstract}

    \maketitle

    \section{Introduction}
        In ecology there is a large interest in the spatial and temporal distribution
        of animals. Depending on the species, the spatial distribution of individuals
        might be independent if they do not interact, clumped if there is some form
        of attraction between them, or evenly spaced for repulsive interaction of
        individuals \cite{Brown1970spacing}. Here, we are especially interested in
        the latter case, more specifically, we are interested in territorial species,
        who inhabit an exclusive territory, which is defended against members of the
        same species. This defense is usually either performed by aggressive behavior
        against intruders, or by deterrent markings of the territory, often by
        auditory signals or olfactory scent marks along the perimeter \cite{Brown1970spacing}.

        Central properties of interest for territories, as well as home ranges, are
        their size. A common method for the determination of the size and
        visualization of home ranges or territories is to calculate the convex hulls
        for the points visited in time for both
        experimental \cite{harris1990homerange,worton1995convex,karki2015estimating}
        as well as simulational data \cite{carter2015modeling}. First studies
        in this directions appeared in the 1940s \cite{mohr1947}. This sparked
        the interest of mathematicians, who started to work on the convex hulls
        of abstract sets of random points, like independently sampled points.
        More interesting and slightly closer to ecology are sets of
        correlated random points. For simple random walks, first the expectation
        value of the perimeter of their convex-hull was studied \cite{spitzer1961}
        and numerous other studies lead eventually to exact results
        for the stochastic properties of the area  \cite{Majumdar2010Random}. Consequently, there is
        quite some interest in the fundamental properties of convex hulls
        \cite{Randon2009convex,Chupeau2015Convex,Grebenkov2017mean}, but exact results
        concerning the full probability distributions
        are missing. Nevertheless, by using numerical large-deviation
        sampling techniques the distribution of perimeter and area of various types of random walks
        could be studied over hundreds of decades in probability
        \cite{Claussen2015Convex,Dewenter2016Convex,schawe2017highdim,schawe2018avoiding,schawe2019true}.

        Although the study of the properties of convex hulls of random points is
        motivated by ecological models, no study of the stochastic properties of territories,
        in particular when addressing the
        full distribution, is known to us where the set of random points originates
        from a more realistic model of the motion of animals. In this work, we
        are treating such a case based on a simple agent-based model
        introduced in Ref.~\cite{Giuggioli2011animal}, where agents perform a random
        walk on a lattice and leave scent marks on visited sites. When encountering
        a foreign scent mark, the agent backtracks away from the adversarial territory.
        This model gives rise to territories and with a slight modification to
        stable home ranges \cite{potts2012territorial}, i.e., the area in which
        an animal usually lives.
        Here, the size of the territory is
        quite straight forwardly defined as the area marked by scent and we will
        study the distribution of this property in very high detail using computer
        simulations \cite{practical_guide2015}. Especially, we will explore the probability density
        function deep into the tails of rare events, which occur with a probability
        of less than $10^{-30}$ and identify the mechanisms
        leading to and the properties of such rare events of individual animals
        with atypically small or large territories. In particular, we will
        make a connection to large deviation theory and characterize the right
        tail of atypically large territories with an approximation of its rate
        function.

        To obtain estimates of the probability density function for the size
        of territories with such a high precision, we need to employ
        Markov chain Monte Carlo importance sampling methods,
        which we will describe after a precise model description in Sec.~\ref{sec:methods}.
        Then, in Sec.~\ref{sec:results}, we will show and interpret the
        results of our simulations. Finally we summarize our findings in
        Sec.~\ref{sec:conclusions}.

    \section{Models and Methods}
    \label{sec:methods}
    \subsection{Model Specification}
        We are studying a model for the emergence of territory by scent marks
        introduced in Ref.~\cite{Giuggioli2011animal}.
        This model lives in discrete time and discrete $d$ dimensional space. There are
        $M$ agents starting on random sites of a lattice with $L^d$ sites and periodic
        boundary conditions, which we call the \emph{world}. At each of the $t=1,\ldots,T$ time
        steps, all agents move synchronously to one of the $2d$ adjacent sites. Each agent marks
        the current site with its individual scent, potentially adding a new scent to already
        existing scents.
        The way the adjacent site is selected, is determined by the scents on the current site at time $t$.
        If there is no scent of other agents on the current site,
        the agent visits at time $t+1$ a uniformly randomly selected adjacent site.
        Otherwise, if there is an adversarial scent, the agent has to step at time $t+1$ on an adjacent site already
        marked with its own scent, i.e., it backtracks into its own territory.
        A scent stays active for $t_\mathrm{a}$ time steps. For $t_\mathrm{a} = 0$ this corresponds
        to non-interacting agents, each performing a standard random walk on a lattice,
        and $t_\mathrm{a} = T$ to static territories. Values in between
        allow the territories to move on a slower time scale as demonstrated in Ref.~\cite{Giuggioli2011animal}.
        In this study, we will concentrate on the static $t_\mathrm{a} = T$ edge case.

        \begin{figure}[htb]
            \centering
            \subfigure[]{\label{fig:examples:2d}
                \includegraphics[width=0.4\linewidth]{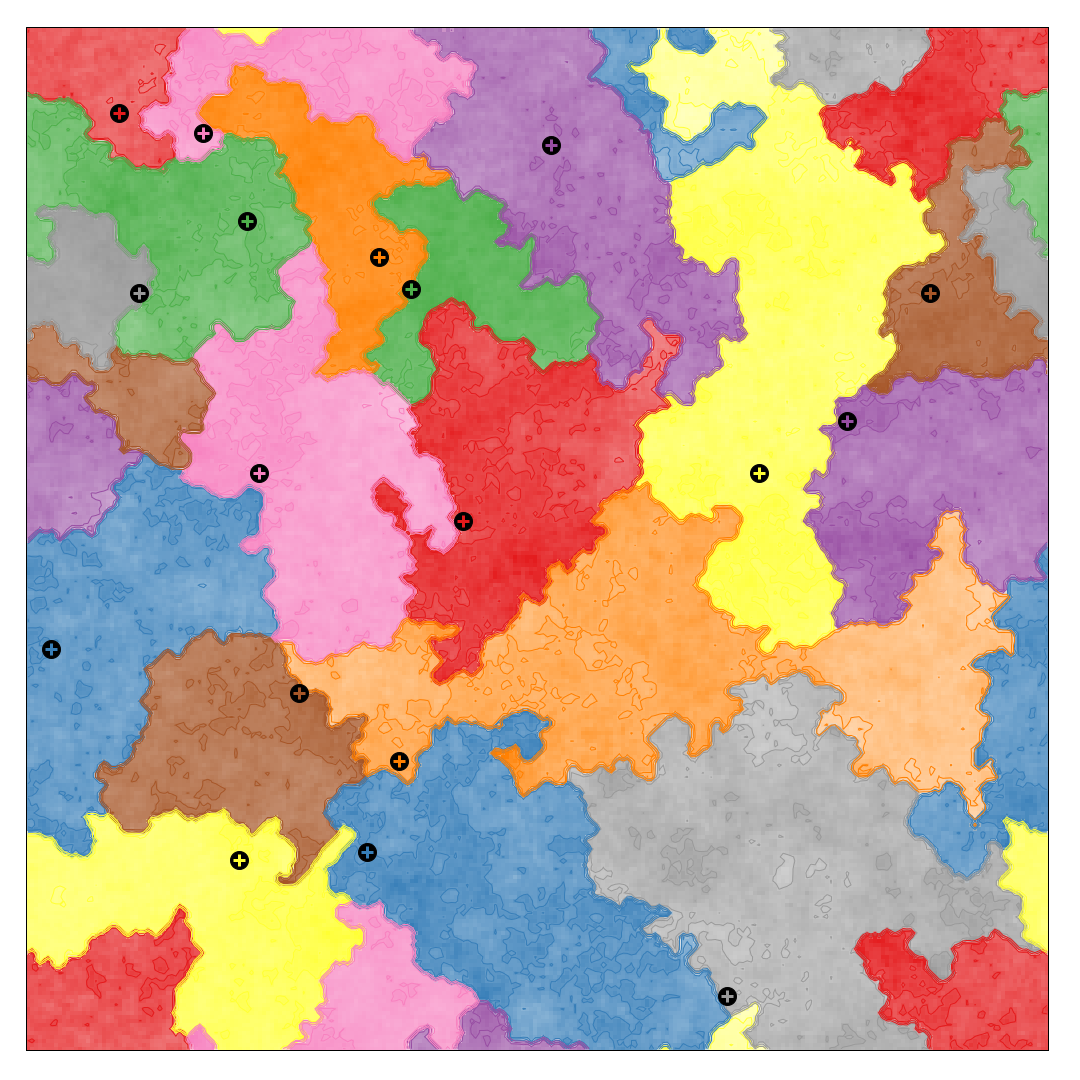}
            }
            \subfigure[]{\label{fig:examples:1d}
                \includegraphics[width=0.44\linewidth]{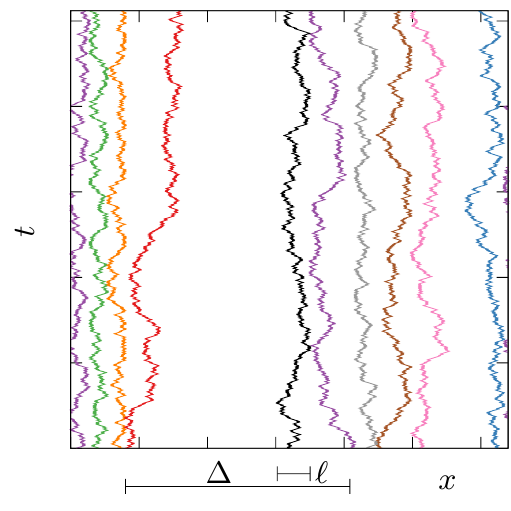}
            }
            \caption{\label{fig:examples}
                Example for the \subref{fig:examples:2d} $d=2$ case of $M=18$ agents
                on a world with periodic boundaries of size $L=256$
                after $T=5 \cdot 10^6$ time steps. The intensity and the contour lines
                show how often a site was visited by an agent, the markers show the initial
                positions. Note that there are only 9 colors and each is assigned to two
                unrelated agents; and \subref{fig:examples:1d} for the $d=1$ case with $M = 10$, $L=320$
                after $T=1024$ steps. Territory of the black agent 0 is marked as the
                distance $\ell$ below the horizontal axis; the distance $\Delta$ is the possible
                space available between the starting positions of agent 0's neighbors.
            }
        \end{figure}

        The interpretation of this model for $d=2$ is quite intuitive, as there are two dimensional
        territories of marked regions arising, if the density of agents is sufficiently large.
        A site is considered as being part of the territory of the first agent who has stepped on it.
        In Fig.~\ref{fig:examples}\subref{fig:examples:2d} an example is shown where each field is colored with an
        agent specific color whose intensity depends on the number of visits to the corresponding
        site (and additional contour lines for clarity). Clearly, the dynamics lead to extremely
        well defined territories.

        However, also the $d=1$ version of this model has direct application, e.g., Ref.~\cite{Giuggioli2011animal}
        uses it to model organisms which actively refresh the scent marks on the perimeter
        of their territory, and compares it with empirical data obtained from foxes.
        The rationale to use the one dimensional version of this model is that a $d=1$ agent
        hitting the border of its territory and refreshing the scent mark on one site of a line
        is similar to a $d=2$ agent walking directly along half of the perimeter of its
        territory---an event which can not be modeled with a random walker in two dimensions.

        Since the main mechanism of this model is the interaction of multiple agents, we have
        to carefully determine the size of the world: too large and there will not be any interaction,
        too small and agents will be restricted to a handful of sites. Especially, we need
        to pay attention how the size of the world should scale with increasing number $M$ of agents
        or larger number $T$ of steps. Since the single agents behave in a diffusive way, we
        scale the size of the world as $L = \lfloor aM\sqrt{T} \rfloor$, where $a$ is a free parameter to determine
        the density of agents. This leads to a roughly similar number of interactions between agents
        when increasing $M$ or $T$, which we checked numerically (not shown). To clarify,
        consider a scaling of the world size
        proportional to $T$, which would fundamentally change the behavior from a crowded world for
        small values of $T$ to free diffusion without any interaction for large values of $T$.

        The observable, we are studying is the total size $\ell_i$ of the territory of agent $i$, i.e.,
        the number of sites marked with the corresponding agent's scent. In the following we will
        mostly concentrate on one arbitrarily chosen ``agent 0'', without loss of generality.
        Its territory will be denoted as $\ell$ without subscript.
        In particular, we are interested in rare territories, which
        are much larger or smaller than typical territories.

    \subsection{Sampling Rare Events}
        To characterize these extremely rare events, we will look at the the \emph{rate function} $\Phi(\ell/T)$,
        which is a central element of \emph{large deviation theory} \cite{Touchette2009large}. It
        describes the behavior of a large class of distribution $P_T(\ell)$ in the limit of, in this case, large times $T$
        and is defined by
        \begin{align}
            P_T(\ell) = e^{-T \Phi(\ell/T) + o(T)}.
            \label{eq:rate}
        \end{align}
        If the distribution $P_T(\ell)$ can be
        described in such a way using a rate function, one says the distribution fulfills a \emph{large deviation principle}.
        Note that $\Phi$ is a function of $\ell/T$, i.e., the ratio of the territory size $\ell$ to the maximum size which
        is possible after $T$ steps.

        Here we want to approximate the rate function $\Phi$ valid for $T\to\infty$ using simulations
        of systems with finite values of $T$. From the distributions of finite sizes $P_T(\ell)$
        we can calculate \emph{empirical} rate functions $\Phi_T$. If they diverge for increasing
        values of $T$, we can exclude the existence of a limiting rate function $\Phi$, but if they
        converge towards such a limiting form, we can even use this to estimate an approximation
        of the functional form of the rate function $\Phi$. In Section~\ref{sec:results} we will
        indeed observe such a convergence for the right tail and observe a functional form very similar to
        standard random walks.

        Obtaining the data of the far tails of the distribution $P(\ell)$, needed to calculate the empirical rate
        functions, is far from trivial. The conventional method of generating independent samples,
        generating a histogram from them and estimate the distribution from that, is limited to
        values of $\ell$ which have a large probability to be observed during a feasible simulation
        time, say, larger than $10^{-10}$. But probabilities of $10^{-30}$, which might be needed to
        characterize the far tails, are far beyond reach of this method.

        To sample events with such low probabilities efficiently, we resort to a Markov chain Monte Carlo (MCMC)
        method, which was used previously for a range of different applications \cite{Hartmann2002Sampling,schawe2019large,chen2019large}
        including the study of areas of convex hulls enclosing the traces of random walks \cite{Claussen2015Convex,Dewenter2016Convex,schawe2018avoiding}. Since the method has been described elsewhere, we only give a brief
        description here, which mainly defines the actual implementation for the present model and few
        general explanations. For our MCMC approach, the \emph{states} of the Markov chains are given by a
        \emph{realization}
        of the set of $M$ random walks. Thus, each state consists of a stochastic simulation itself,
        of the actual random walks, and the random walks are embedded into a higher-level Markov chain.

        First, as for any Markov chain method with the Metropolis-Hastings algorithm \cite{newman1999monte},
        we have to define a \emph{change move} to generate trial realizations. While
        there are elaborate and efficient change moves for the simple random walk or the
        self-avoiding random walk used for polymer simulations \cite{Madras2013},
        we are not aware of any prior work
        for the mutually-avoiding random walks of the territoriality model we study.
        The growth mechanism of the territoriality model does indeed prevent the use of
        methods similar to the ones used for the mentioned non-growing random walk models,
        (which is explained in a bit more detail in \cite{schawe2018avoiding}).
        Therefore we resort to a method which does not operate on the random walk itself,
        but on the random numbers used by the computer program to generate the random walk.
        This method was introduced to study non-equilibrium processes in Ref.~\cite{Hartmann2014high}
        and successfully applied to different models defined by growth
        processes \cite{schawe2018avoiding,schawe2019true,schawe2019large}. To understand the approach,
        note that for any stochastic simulation it is necessary to construct a realization of
        the studied ensemble from a sequence of random numbers. Clearly, it does not change the behavior
        if one first generates all the
        random numbers, stores them in a vector, and uses them for the actual stochastic simulation.
        Thus, instead of constructing
        the Markov chain from realizations and propose change moves applied to the realizations,
        we build a Markov chain consisting of random number vectors as states and apply change moves
        to these vectors. This requires a lot of computational power, because generally after each change
        move a new realization of $M$ random walks has to be constructed from a vector of random numbers.
        But this approach is quite generally applicable to models, for which specialized change moves on the
        realizations are not trivial to construct.

        For the Metropolis-Hastings algorithm the change moves will either be accepted, i.e., used as the next state
        of the Markov process, or rejected,
        i.e., the current state is kept, according to the Metropolis acceptance probability
        $p_\mathrm{acc} = \min \left\{ 1, e^{-\Delta E / \Theta} \right\}$ \cite{metropolis1953equation}.
        Here $\Theta$ is an artificial temperature, whose role will be explained in the next paragraph.
        We identify the ``energy'' $E$ with our observable of interest, which is obtained
        from each realization, here $E \equiv \ell$, and $\Delta E$
        denotes the change in this quantity between the current realization and the proposed
        trial realization.
        This way, realizations $\mathcal{C}$ will eventually be distributed according
        to the Boltzmann distribution
        \begin{align}
            Q_\Theta(\mathcal{C}) = \frac{e^{-\ell / \Theta}}{Z(\Theta)} Q(\mathcal{C}),
            \label{eq:q}
        \end{align}
        where $Q(\mathcal{C})$ is the natural, unbiased distribution and $Z(\Theta)$ the partition function,
        which just takes the role of a normalization constant for our purposes.
        Conveniently, we can calculate the distribution $P(\ell)$ from the distribution
        of realizations by summing the probabilities of all realizations with the same value of $\ell$
        \begin{align}
            P(\ell) = \sum_{\{\mathcal{C} | \ell(\mathcal{C}) = \ell\}} Q(\mathcal{C}),
        \end{align}
        and analogously for the distribution $P_\Theta(\ell)$ measured in the biased Markov chain.

        We can choose the artificial ``temperature'' $\Theta$ freely, which allows us due to
        Eq.~\eqref{eq:q} to tune the typical values of $\ell$ encountered in the Markov chain
        and therefore guide the Markov chain into atypical ``energy'' ranges with a careful choice
        of $\Theta$.
        Most importantly for our application, the bias is well defined in Eq.~\eqref{eq:q}
        and can be removed \cite{Hartmann2002Sampling} with the knowledge of $Z(\Theta)$. We can obtain this by exploiting the
        uniqueness of $P(\ell)$, i.e., if we have estimates for two biased distributions at
        different values of the artificial temperature $P_{\Theta_i}(\ell)$ and $P_{\Theta_j}(\ell)$,
        we can calculate the ratio of the two corresponding $Z(\Theta_i)$ via
        \begin{align}
            e^{\ell / \Theta_i} Z(\Theta_i) P_{\Theta_i}(\ell) = e^{\ell / \Theta_j} Z(\Theta_j) P_{\Theta_j}(\ell).
        \end{align}
        This requires us to obtain estimates for biased distributions at multiple values
        of the artificial temperature, such that they overlap pairwise. Also the
        statistical precision of the estimate in the overlapping region should be
        decent to avoid large statistical uncertainties.

        As for all Markov chain
        Monte Carlo techniques, the subsequent realizations in the Markov chain
        are correlated. So one has to ensure that the Markov process is in equilibrium
        before taking measurements and to perform enough change moves before
        taking the next measurements to allow the samples to decorrelate \cite{newman1999monte}.
        For the results shown in the
        following section, we used about a dozen temperatures for each
        parameter set to obtain estimates over a very large part of the support
        of $P(\ell)$. Note that the simulations for different artificial temperatures
        are completely independent and can be performed in parallel.

        This sampling method also allows us to measure other properties of
        the encountered samples. While a full joint probability distribution would
        need two independent artificial temperatures and much more simulation time,
        we can use the values of a second observable $o$ encountered during one
        simulation, to construct a partial joint probability distribution using
        Bayes' theorem. Because $\ell$ is biased, the pairs $(o, \ell)$ encountered
        during the simulation can not be used to estimate the joint probability $P(o, \ell)$
        directly, but they can be used to estimate the conditioned
        probability $P(o | \ell)$. After the simulation we obtain $P(\ell)$ from the
        data, as described above. Now Bayes' theorem allows us to obtain a part of the
        joint probability density $P(o,\ell) = P(o | \ell) P(\ell)$.

    \section{Results}
        \label{sec:results}

        Markov chain Monte Carlo methods are still rather compute intensive and a
        systematic study of the two-dimensional territories is infeasible, as we explain now: It necessitates a
        number of agents $M$
        which increases quadratically in the linear size of the system to preserve a fixed density.
        The linear size of
        the system must be larger than $T$ to avoid extreme walks to interact with themselves (or
        for non-periodic borders with the boundary). At the same time the interaction between the
        different agents leads to a problem when choosing the change moves: A single changed step
        of a single agent will cascade via the interactions through the whole system and often
        introduce a substantial change, which has to be rejected. This is aggravated by the fact
        that quite large times $T$ are necessary to observe the formation of territories.
        Since current studies of the full distribution of home range areas for
        comparable random walk models with a single random walker are limited to
        $T < 10000$ \cite{Claussen2015Convex,Dewenter2016Convex,schawe2017highdim,schawe2018avoiding},
        even though there exist efficient change moves for those models, the two-dimensional
        case of the territory model is beyond reach at the moment.

        Fortunately, the problems of the two-dimensional case either vanish or are far less severe
        in the one-dimensional case. In one dimension
        agents can not go beyond the initial positions of their direct neighbors, such that the
        linear size may be smaller than $T$ without risking self-interaction. This also reduces
        the severity of the cascading of change moves through the whole systems, since mostly
        only the two direct neighbors will be affected. Finally, the number of agents $M$ scales only
        linear in the linear size of the system to preserve a fixed density, instead of quadratically.
        Therefore, we will present in the following sections our
        numerical results characterizing the one-dimensional case and compare them to a standard
        random walk on the same lattice (the $M=1$ case).

        Before we dive into the full distribution, it is useful to take a look
        at the behavior of the mean territory size $\avg{\ell}$. This can be
        obtained via simple sampling, such that we have access to larger systems
        than for the study of the full distribution.

        One of the fundamental properties of simple random walks is their diffusive
        behavior, i.e., observables characterizing their size along one dimension
        scale as $\sqrt{T}$ in the number of steps, i.e., $\avg{\ell} / \sqrt{T} \to \mu$,
        for $T\to\infty$ where $\mu$ is a constant. We expect this for the present model as well,
        which is reflected by the scaling of the world size like $\sqrt{T}$. Our results
        shown below are consistent with that.
        Also the value of $\mu$ and its dependence on $a$ or $M$ are still of interest.

\begin{figure}[htb]
            \centering
            \includegraphics[scale=1]{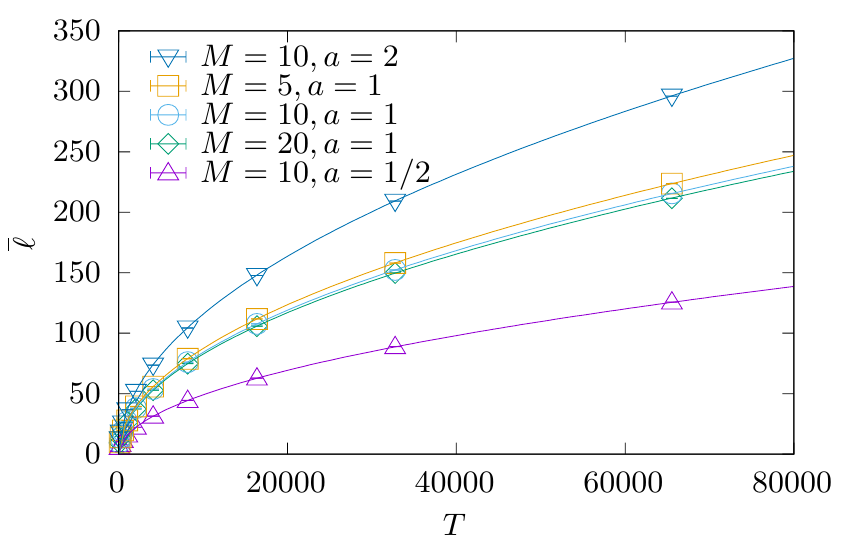}
            \caption{\label{fig:means}
                Behavior of the mean territory size for a selection of parameters $M$ and $a$.
                Fits are to $\ell = \mu_{M,a} \sqrt{T} + C_1$ for $\ell \ge 1024$. The resulting values
                are $\mu_{5, 1} = 0.8737(3)$, $\mu_{10, 1/2} = 0.4905(2)$, $\mu_{10, 1} = 0.8418(3)$, $\mu_{10, 2} = 1.1572(3)$, and $\mu_{20, 1} = 0.8272(3)$.
            }
        \end{figure}

        Therefore we show in Fig.~\ref{fig:means} the average territory size $\overline{\ell}$
        measured over $10^6$ random realizations for different numbers of steps taken $T$ and
        different values of the parameters determining the number $M$ and density $a$
        of agents. Here we choose intermediate values of $a$, since very large ones would
        inhibit the interaction between the agents. The solid lines are fits to the
        form $\ell = \mu \sqrt{T} + C_1$,
        where the first term mirrors the dominating diffusive scaling behavior and the
        second term should account for corrections to this scaling for finite sizes.

        Indeed, the fits of this form describe the behavior well (with $\chi^2_\mathrm{red}$
        goodness of fit values between $0.7$ and $1.2$ for all shown cases.) The fit results for $C_1$ are always
        smaller than $0.5$, i.e., they have no visual impact on Fig.~\ref{fig:means}.
        The values of $\mu$
        obtained by the fits are listed in the caption of Fig.~\ref{fig:means} and are
        much smaller than the known value of the \emph{span} of simple random walks. The span is the distance from the
        leftmost to rightmost visited point, which is analogue to our territory. For a one-dimensional
        lattice with unit spacing it is known to be $\mu = \sqrt{8/\pi} = 1.596..$ \cite{daniels1941probability},
        i.e., the interaction of the agents has a large influence on the \emph{typical} behavior.

        Interestingly the choice of $M$, despite not changing the density of the agents,
        has an influence on the asymptotic territory size, as the value of $\mu$ decreases with growing value of
        $M$.
        Since we do not look at any special agent, this must mean that more agents lead to
        a higher proportion of unclaimed territory. Less surprising is that a decreased
        density (larger values of $a$) of agents does not lead to proportionally more
        territory. Especially for large values of $a$, the limiting factor is not the
        amount of unclaimed territory, but the diffusive behavior of the agents which have
        access to the largest area of free territory, while other agents are restricted to small areas
        due to unfavorable initial conditions. This effect, however should diminish greatly
        in the two-dimensional version of the model.

\begin{figure}[htb]
            \centering
            \includegraphics[scale=1]{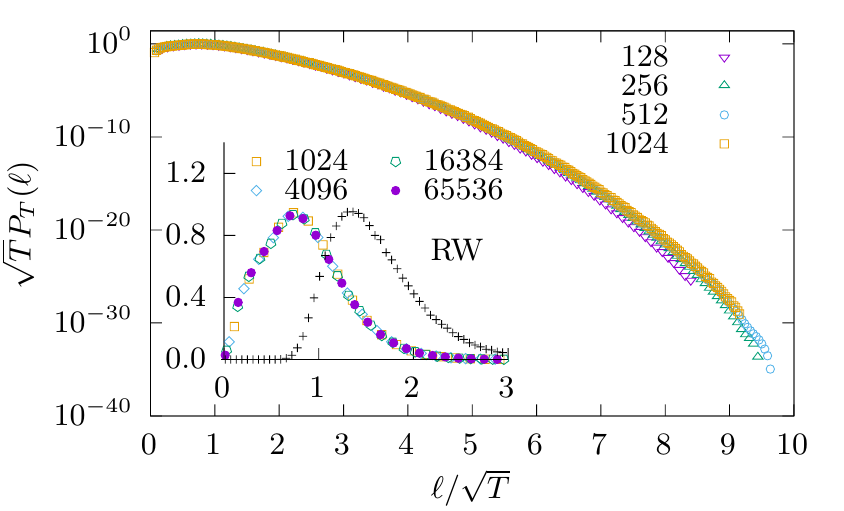}
            \caption{\label{fig:dist}
                Distribution $P_T(\ell)$ for different system sizes $T \in \{128, 256, 512, 1024\}$
                including very rare configurations. The axes are scaled to collapse all sizes on
                a size-independent scaling form.
                The inset shows the high probability part with
                a linear instead of logarithmical axis with data obtained via simple sampling
                and therefore larger values of $T \in \{1024, 4096, 16384, 65536\}$.
                Also shown is a standard random walk (RW) for comparison.
                Not all available data points are shown for clarity.
            }
        \end{figure}

        For standard random walks without territorial competition it is well known that not only the
        means but the whole distribution of the span shows a $T$-independent form
        when scaled with $\sqrt{T}$ \cite{hughes1996random,Kundu2013exact}.
        If this scaling is also valid for the territoriality model under scrutiny, we
        would expect that the distributions measured for different values of $T$ would
        collapse on the same $T$-independent scaling form, i.e.,
        \begin{align}
             \widetilde{P}(\ell / \sqrt{T}) = \sqrt{T} P_T(\ell).
            \label{eq:scaling}
        \end{align}
        In the inset of Fig.~\ref{fig:dist} we can see that this collapse works well in
        the high-probability region. Also there is a comparison to the distribution $P(\ell)$
        of a simple random walk on a lattice, which has its maximum at larger values of $\ell/\sqrt{T}$,
        which is consistent with our results for the mean values, from above.
        In the main plot of Fig.~\ref{fig:dist}, we can observe the same effect over almost the
        whole distribution. Only in the far right tail,
        where effects of the finite-size world come into play, deviations from the common curve
        are strong.

\begin{figure}[htb]
            \centering
            \subfigure[]{\label{fig:examples:small}
                \includegraphics[width=0.44\linewidth]{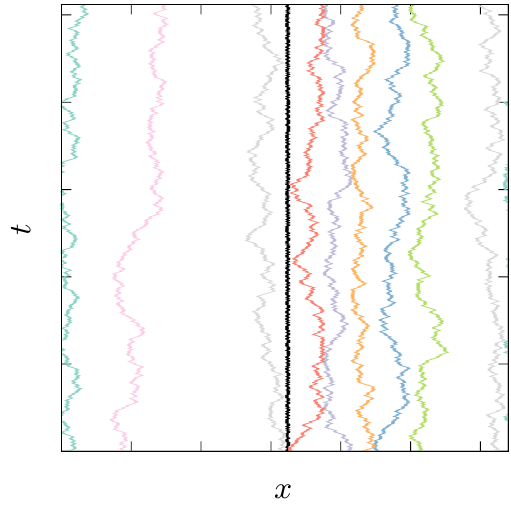}
            }
            \subfigure[]{\label{fig:examples:large}
                \includegraphics[width=0.44\linewidth]{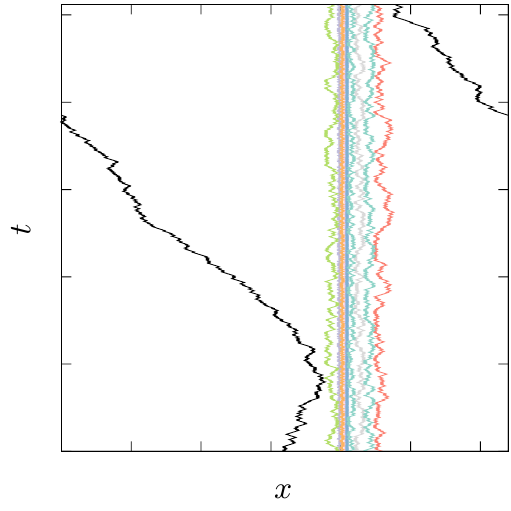}
            }
            \caption{\label{fig:examples:biased}
                Example configurations with $T=1024$, $L=320$, $M=10$ obtained where
                agent 0 (black lines) is biased to
                \subref{fig:examples:small} small territories and
                \subref{fig:examples:large} large territories. Apparently a large
                contribution to the territory an agent can annex, depends on the
                initial positions, either confining the agent, resulting in very small
                territories, or confining most other agents, resulting in a very large
                territory.
            }
        \end{figure}

        To study the far tails of extremely rare configurations in more detail, it is useful to
        look at examples of such rare instances. In Fig.~\ref{fig:examples:biased} examples
        for both the left and right tail are shown. They are realizations from the
        equilibrium distribution of the artificial temperature ensemble at $\Theta = 1$
        and $\Theta = -1$ respectively.

        First, consider Fig.~\ref{fig:examples:biased}\subref{fig:examples:small} corresponding to
        a realization from the left tail. Apparently agent 0, marked in black, is confined from the
        very beginning by its two neighbors which start extremely close to each other. All other
        agents seem to behave quite typically. If we assume that this trapping mechanism is
        the dominant mechanism leading to very small territory sizes $\ell$, we can try to estimate the behavior
        of the left tail of the distribution.
        Therefore we need to know the distribution of the distance $\Delta$ from the left to
        the right neighbor of agent 0.
        Since the starting positions are almost independent (the only correlation arises by
        the impossibility of two walkers starting on the same site), we can approximate the initial positions
        as a Poisson point process. Note that the Poisson point process lives on a continuum, while we
        have a discrete lattice. Therefore the approximation becomes better for increasing size of the
        world $L$. Fortunately, for a Poisson point process with a point density of, here, $\lambda = M/L$
        the distribution of the size $\delta$ of Voronoi cells, i.e., half the distance $\Delta=2\delta$
        to the left and right nearest neighbors, is known
        to be $P(\delta) = 4\lambda^2 \delta e^{-2\lambda\delta}$ \cite{ferenc2007size}.

\begin{figure}[htb]
            \centering
            \includegraphics[scale=1]{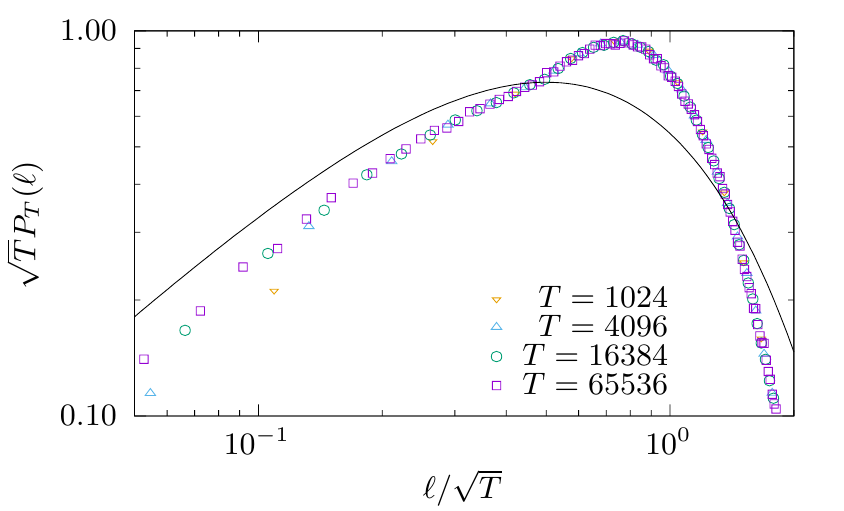}
            \caption{\label{fig:rate_n_left}
                Rescaled probability densities $P_T(\ell)$ in a log-log plot to emphasize the left tail
                for walks with different number of steps $T$. The solid line
                is the distribution \eqref{eq:lefttail}. The size of the world
                scales as $L = \lfloor aM\sqrt{T} \rfloor$ with $M = 10$ agents and scale factor $a = 1$.
            }
        \end{figure}

        For small values of $\Delta$ the diffusion of the agent would cover all of the
        available area. Due to the competition of its adversaries
        we would expect that for a given (small) value of $\Delta$ an area of
        $\Delta/2 = \delta$ would be claimed by agent 0 on average.
        Thus we use $\delta \approx \ell$ and expect
        \begin{align}
            P_T(\ell) \approx 4 \lambda^2 \ell e^{-2\lambda\ell}
            \label{eq:lefttail}
        \end{align}
        for small values of $\ell$.
        We compare this approximation to the data we simulated in double logarithmic axes to
        emphasize the left tail. We rescale the axes the same way as in Fig.~\ref{fig:dist},
        to enable the visualization of very different sizes in the same plot. Also, this scaling
        lets Eq.~\eqref{eq:lefttail} collapse on a $T$-independent scaling form.
        Note that the events in the left tail are
        probable enough that we can observe the whole tail using our simple sampling results: they often
        reach the very leftmost point $\ell = 2$ of minimal possible territory in this model.
        First, we can see in Fig.~\ref{fig:rate_n_left} a reasonable matching, with a deviation
        of $20\%$ over most of the tail, between our measurements and the approximation.
        The matching becomes better for larger values of $T$, which is expected,
        since the approximation Eq.~\eqref{eq:lefttail} becomes better in that case.
        Thus, the initial positions of the agents determine the left tail of the distribution of
        territories in a major way. We will explore this in more detail using correlations
        of multiple observables later in this manuscript.

        \begin{figure}[htb]
            \centering
            \includegraphics[scale=1]{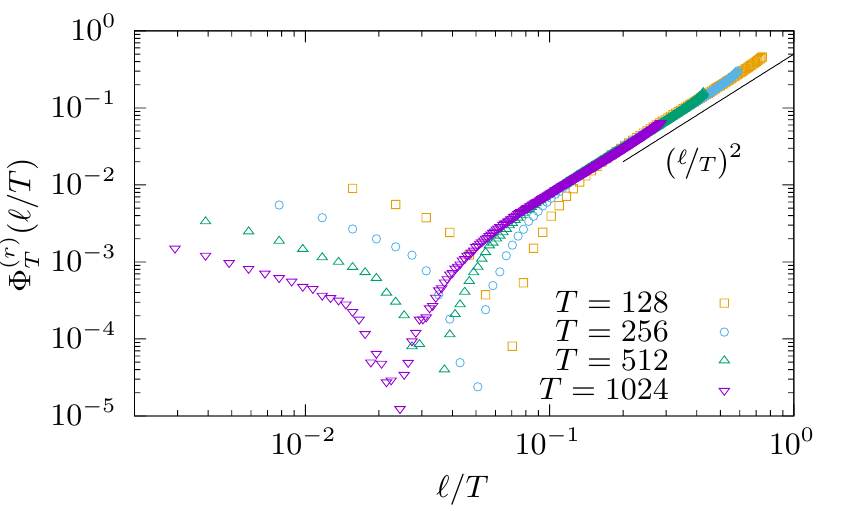}
            \caption{\label{fig:rate_n}
                Empirical rate functions in a log-log plot to emphasize the power-law behavior
                of the right tail for walks with different number of steps $T$. The size of the world
                scales as $L = \lfloor aM\sqrt{T} \rfloor$ with $M = 10$ agents and scale factor $a = 1$.
            }
        \end{figure}

        For the right tail of larger than typical territories the instances seem to consist of extremely
        dense initial conditions for all agents, in such a way that the agent with the largest territory can occupy almost
        the whole world, which is shown in Fig.~\ref{fig:examples:biased}\subref{fig:examples:large}.
        The total size of the territory should
        therefore depend somewhat on the size of the world. However, the ``straight line'' movement in the
        example suggests a ballistic character of the agents with extremely large territories.
        In this case $\ell$ should scale like the number of steps $T$ in the far right tail for
        worlds which are large enough. Therefore,
        we scale the horizontal axis of the rate function like $\ell/T$ in Fig.~\ref{fig:rate_n}, i.e.,
        \begin{align}
            P_T(\ell) = e^{-T \Phi^{(r)}(\ell / T)}.
            \label{eq:rate_scaled}
        \end{align}
        Further we subtract the minimum, i.e., shift all empirical rate functions such
        that their unique minimum has a value of zero, which is a property
        of the rate function. This does therefore not change the form to which they
        converge.
        Indeed, for our data we observe that the right tails converge to a common
        form, which behaves like $\Phi^{(r)} \propto {\left(\ell/T\right)}^2$ until
        effects of the limited world size and therefore limited
        territory size truncate the distribution. This is the same behavior of the rate function
        as for a single random walk observed before \cite{Claussen2015Convex,schawe2017highdim}.
        Assuming the rate function behaves like a power law for large values of $T$,
        $\Phi^{(r)}_T(\ell / T) \to \left(\ell/T\right)^\kappa$, one can understand the value of the
        exponent $\kappa = 2$ by comparing the form of the distribution expressed via
        the rate function in Eq.~\eqref{eq:rate_scaled} with the
        scaling Eq.~\eqref{eq:scaling} demonstrated in Fig.~\ref{fig:dist}.
        Since this should only be valid for large values of $T$, we can neglect the $\sqrt{T}$
        factor in Eq.~\eqref{eq:scaling} and arrive at
        \begin{align}
            \exp\left(-T\Phi^{(r)}(\ell/T)\right) \approx \widetilde{P}\left(\ell/\sqrt{T}\right)
        \end{align}
        Since $\widetilde{P}$ has no explicit dependency on $T$, we must be able to formulate
        the left hand side as a function of the same argument $\ell/\sqrt{T}$:
        \begin{align}
            \exp\left(-T\Phi^{(r)}(\ell/T)\right) &\approx \exp\left(-T (\ell/T)^\kappa \right)\\
                                         &= \exp\left(- {\left(\ell / T^{(\kappa-1)/\kappa}\right)}^\kappa \right)
        \end{align}
        and therefore $(\kappa-1)/\kappa = 1/2 \Rightarrow \kappa = 2$.
        Note that this argument is more generally stated in Ref.~\cite{schawe2018avoiding}.
        One can observe a collapse onto a common curve
        for the right tail onto this form, which suggests that the large deviation principle holds here
        and no fundamental differences to standard random walks exist for the limit of extremely large
        territories---though the detailed shape and location of the typical region differ a lot
        (cf.~Fig.~\ref{fig:dist}).

        In the almost complete probability density functions we showed, it is obvious that
        very large territory sizes are far more rare than very small territory sizes.
        This can be made plausible by the following simple argument: While left-tail events
        only need two arbitrary agents to start close to agent 0, right tail events need
        every agent to start in a very small region. Also while left tail events do not need
        any rare configuration of steps, since the starting positions are already sufficient
        to restrict agent 0 to a very small territory, right tail events need a rare configuration
        of steps from agent 0, to span the available territory. For extremely right-tail
        events, even the neighbors need to show rare subdiffusive behavior to not claim
        territory before agent 0 arrives.

\begin{figure}[htb]
            \centering
            \subfigure[]{\label{fig:cmp:a}
                \includegraphics[scale=1]{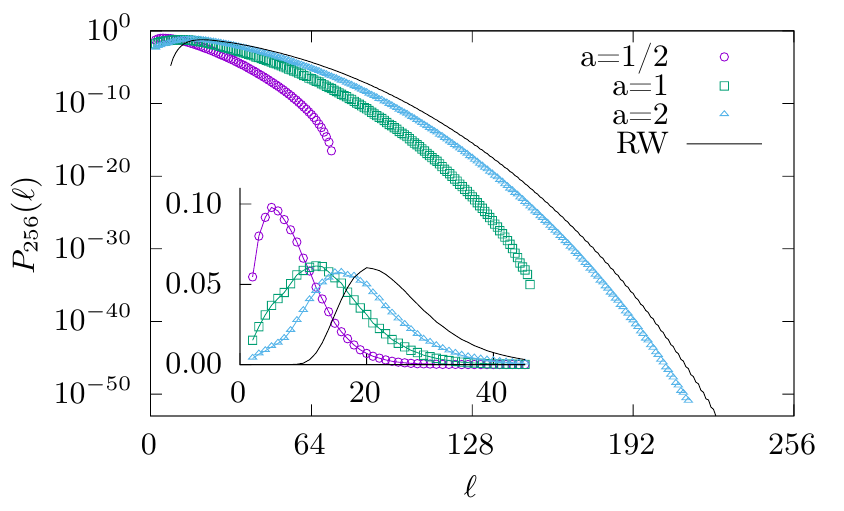}
            }
            \subfigure[]{\label{fig:cmp:m}
                \includegraphics[scale=1]{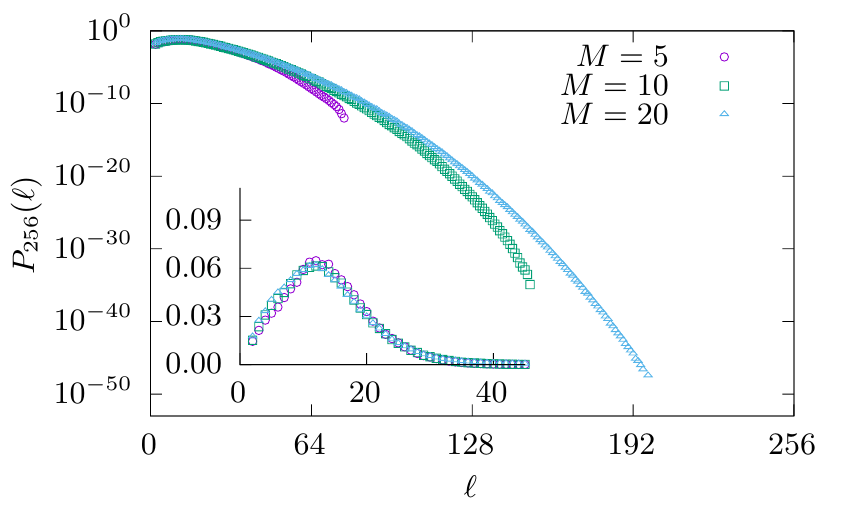}
            }
            \caption{\label{fig:cmp}
                Parameter study. $L = \lfloor aM\sqrt{T} \rfloor$.
                \subref{fig:cmp:a} varying $a$ ($M=10, T=256$), lines visualize a standard random walk (RW), \subref{fig:cmp:m} varying $M$ ($a=1, T=256$).
                The $M=10, a=1$ data set is the same as shown in Figs.~\ref{fig:dist} and \ref{fig:rate_n} for $T=256$.
            }
        \end{figure}

        Although, we limited this study to the static limit $t_a = T$, there are still free parameters like the
        density $a$ and number of agents $M$. We will test their influence on the full distribution
        with a very short parameter study shown in Fig.~\ref{fig:cmp}.

        First, we should consider the influence we expect for different values of $a$: Since $a$ directly
        governs the density of agents due to $L \propto a$, one would expect for small values of $a$,
        i.e., high densities,
        that the distribution $P(\ell)$ becomes concentrated around the minimum $\ell = 2$. For large
        values of $a$, the density decreases so far as to prohibit interactions between agents,
        such that $\ell$ will behave the same way as the span of a single random walk. Both expectations
        are met when looking at the inset of Fig.~\ref{fig:cmp}\subref{fig:cmp:a}. Here for
        smaller values of $a$ a more pronounced peak at lower values of $\ell$ arises and
        the curve for larger values of $a$ becomes more similar to the curve of
        a single random walk shown as a black line. Moreover, the main plot shows
        the behavior of the tails. Here, a truncation effect becomes visible, as
        the world size $L$ becomes smaller than $T$ and the extremely large territories do not
        fit into the world anymore.

        Next, consider the influence of $M$. Since the density of agents
        is independent of $M$, due to the scaling of $L \propto M$, we do not expect
        a large influence of $M$ on the typical regime beyond the slight influence we already
        observed for the mean value in Fig.~\ref{fig:means}. While very low values of $M$, like $M = 2$
        will surely impact the interaction between the walkers, this effect should diminish quickly
        for larger values. Indeed, the distribution $P(\ell)$ for $M \ge 5$, which we visualized in
        the inset of Fig.~\ref{fig:cmp}\subref{fig:cmp:m}, are very similar to each other.
        However, for the far tails visualized in the
        main plot, considerable differences become apparent.
        Since $L \propto M$, we encounter the same truncation as visible for small $a$.
        So larger values of $M$ allow us to explore deeper into the right tail---unfortunately
        the computational cost also increases with $M$.

\begin{figure}[htb]
            \centering
            \includegraphics[scale=1]{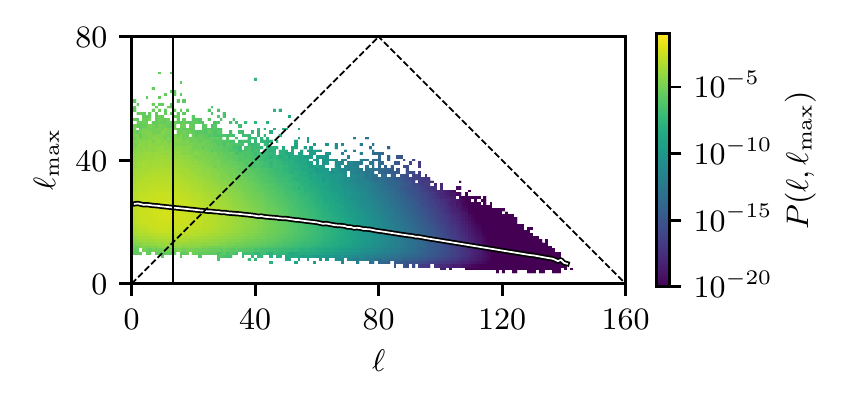}
            \caption{\label{fig:bayesian2d}
                Joint probability distribution of $P(\ell, \ell_{\max})$
                showing correlation of the size of the territory of agent 0 with the largest
                territory of its adversaries for $T=256$, $M=10$, $a=1$, $L=160$. The dashed lines
                mark $\ell_{\max}= \ell $ (rising), $\ell_{\max}+\ell=T$ (descending). The vertical line
                shows the typical value of $\ell$ and the white line with black outline the average
                of $\ell_{\max}$ restricted to the given value of $\ell$.
            }
        \end{figure}

        Previously, we considered a few extreme example configurations to get a feeling for the structure
        of extreme configurations. To get a more complete and quantitative picture, we can instead scrutinize
        the joint probability of two characteristic observables. From the data collected during the necessary
        simulations for determining $P(\ell)$, using Bayes' theorem as described in Sec.~\ref{sec:methods},
        we can determine very large parts of the joint probabilities of $\ell$ and any observable,
        as shown in Fig.~\ref{fig:bayesian2d}, for very little additional computational cost.

        First, we show in Fig.~\ref{fig:bayesian2d} the joint probability
        of $\ell$ and the size of the largest territory of all other agents $\ell_{\max} :=
        \max_{i>0} \ell_i$. Parts of the joint
        distribution are marked white, if there are no data. Note that the upper right triangle above
        the dashed line has a probability of $0$, since the sum of the two territories must be smaller or equal
        to the size of the world, also the $M-2$ other agents block at least a small territory.
        The white line with black outline shows the mean values $\overline{\ell_{\max}}(\ell)$ for each
        value of $\ell$, it exhibits an expected slight anti-correlation. This anti-correlation is due
        to the fact that all agents share the same world, if one occupies more, the others get less.
        The distribution can be split
        into two parts, one where the territory of agent 0 is dominating, below the line given by
        $\ell_{\max}= \ell$, and another where one of the other agents covers the largest territory, above this line.
        In the region where agent 0 has a territory of typical size, marked by the vertical line, almost always
        at least one other agent covers a larger territory. The distribution is located in the figure
        above the rising dashed line.
        In particular one can learn from this figure that only when agent 0 covers about twice of its typical
        territory, where the rising dashed line crosses the $\overline{\ell_{\max}}(\ell)$ line near $\ell=25$,
        agent 0 covers
        among all agents the largest part of the world. This is where the real large-deviation behavior sets in.

\begin{figure}[htb]
            \centering
            \subfigure[]{\label{fig:bayesian2d_eff:start}
                \includegraphics[scale=1]{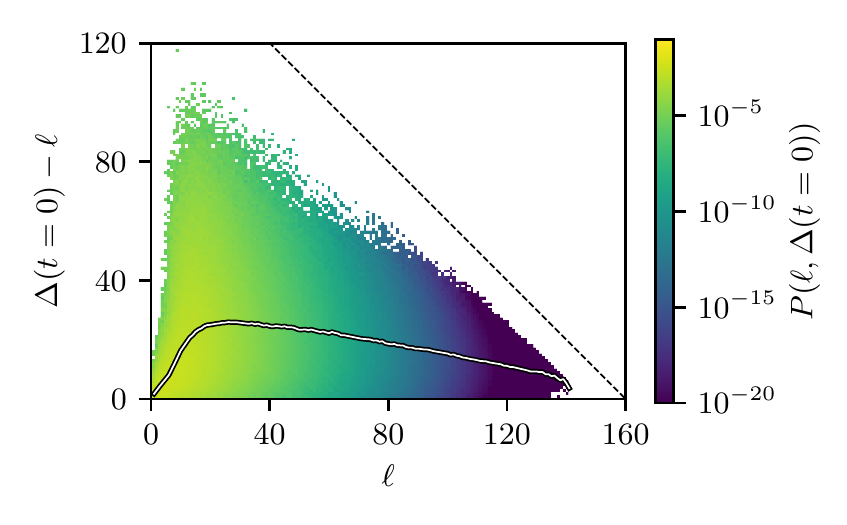}
            }
            \subfigure[]{\label{fig:bayesian2d_eff:end}
                \includegraphics[scale=1]{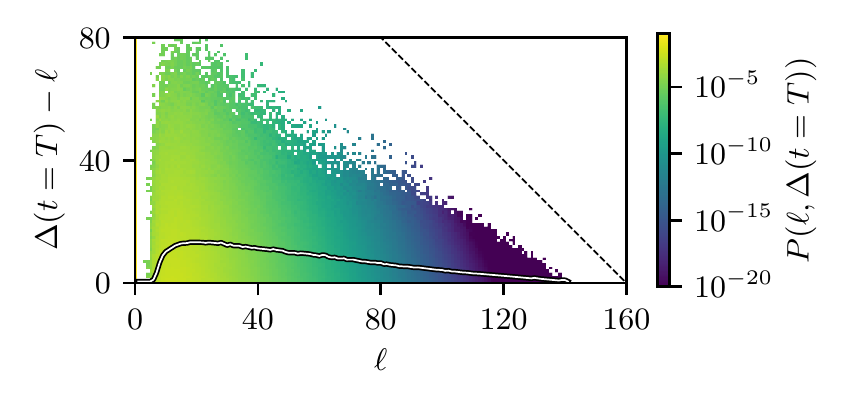}
            }
            \caption{\label{fig:bayesian2d_eff}
                Joint probability distribution of $P(\ell, \Delta)$ showing correlation of the
                size of the territory of agent 0 with \subref{fig:bayesian2d_eff:start}
                the distance $\Delta(t=0)$ between its initial neighbors
                and \subref{fig:bayesian2d_eff:end} the distance between the final territories of
                its neighbors $\Delta(t=T)$ for $T=256$, $M=10$, $a=1$, $L=160$.
            }
        \end{figure}

        Next,  in Fig.~\ref{fig:bayesian2d_eff}, we want to study the efficiency of the agents.
        Here again we mark the inaccessible region in the top right of the diagram by a dashed line.
        Also here, the conditioned mean at a given value of $\ell$ is indicated by a white line with black border.

        First, in Fig.~\ref{fig:bayesian2d_eff}\subref{fig:bayesian2d_eff:start} we look at
        the observable $\Delta(t=0) - \ell$, which is how many sites, of those which are
        accessible given the starting positions, are not claimed. This is a measure of inefficiency. Since lower values
        of $\Delta(t=0) - \ell$ signify a more efficient use of the available space,
        we encounter the---only at the first sight---counter-intuitive result, that agents with exceptionally small
        territories are still exceptionally efficient. This can be explained by the fact
        that the initial positions are indeed the driving factor resulting in very small
        territories. Also note that at very small values of $\ell$ we encountered no
        realizations with even typical values of $\Delta$, shown by the large white area for
        small values of $\ell$. Also we note that for
        small values of $\ell$ the mean of $\Delta(t=0) - \ell$ has a slope of almost one,
        which supports our guess that in this case the agent should claim on average half
        of the initially available space.
        The maximum of this inefficiency measure is reached around the
        typical realizations of $\ell \approx 0.84\cdot \sqrt{256} \approx 13.4$.
        To reach larger than typical territory sizes, the inefficiency has to decrease
        again, since agents need to claim larger portions of the limited total size.

        Also we study $\Delta(t=T) - \ell$, which is how many sites are still available to
        agent 0 in the end of the simulation. This is a measure to estimate how large the role
        of confinement by their neighbors is. Small values indicate that the agents were limited
        by scent marks left by their neighbors, while large values indicate that there was still
        much territory unclaimed and the agent was limited by its diffusive character.
        In accordance with our observation that agents with smaller than typical territories are constrained by
        the small territory available, Fig.~\ref{fig:bayesian2d_eff}\subref{fig:bayesian2d_eff:end}
        shows that almost all agents in this category claim every last site. Similarly,
        agents with extremely large territories obtain them by not leaving sites unclaimed.
        Only in the region of typical instances we see realizations which leave significant portions
        of sites unclaimed. Overall, there is a high similarity of
        Fig.~\ref{fig:bayesian2d_eff}\subref{fig:bayesian2d_eff:start} and
        Fig.~\ref{fig:bayesian2d_eff}\subref{fig:bayesian2d_eff:end}, which shows that rare realizations
        are mostly determined by rare initial positions of the agents, rather than by rare spatio-temporal
        evolutions.

    \section{Summary and Outlook}
    \label{sec:conclusions}
        We studied a model for the emergence of territories by scent marks left by random
        walkers in one-dimension, which was used before to model the territorial behavior
        of foxes \cite{Giuggioli2011animal}. The typical, i.e., high-probability, behavior of our model turns
        out to be very different from the typical behavior of standard random walks.
        Using sophisticated large-deviation
        Markov chain Monte Carlo algorithms,
        we are able to obtain the distribution of the territory size over almost the full range
        of the support and many decades in probability. For the analysis,
        we concentrate on the behavior of extreme
        realizations, in which one individual either claims an extremely large or extremely
        small territory. The results indicate
        that the far right tail can be described by a rate function of a power
        law shape with exponent $2$. This is similar to the behavior found for non-interacting
        random walks, which show that the presence of interactions, which lead to the
        emergence of territories, does not change the large-deviation behavior substantially. This is
        also a good justification for the previous studies motivated by complex interacting systems
        to scrutinize the rare-event properties of simple models.
        Further, we use large parts of the joint probability densities of the territory size
        and the size of the largest adversarial territory or the size of unmarked territory,
        as well as examples of very atypical territory realizations to gain insight into
        the processes leading to atypical realizations.

        For further studies in the rare-event range,
        it would probably not be very interesting to study the behavior as a function
        of the lifetime $t_a$ of the scent. The reason is that for the present study with $t_a=\infty$ we already found
        an asymptotic similarity of these repelling, but not self-avoiding, random walks with
        the pure random walk model. Nevertheless,
        it could be interesting to see whether this similarity still exists for larger dimensions than one.
        But for that purpose
        a much higher numerical effort would be needed.

    \section*{Acknowledgments}
        HS acknowledges the grant HA 3169/8-1 of the German Science Foundation (DFG) and
        Labex MME-DII (Grant No. ANR reference 11-LABEX-0023).
        The simulations were partially performed at the HPC Cluster CARL, located at
        the University of Oldenburg (Germany) and funded by the DFG through
        its Major Research Instrumentation Programme (INST 184/108-1 FUGG)
        and the Ministry of Science and Culture (MWK) of the Lower Saxony
        State.

    \bibliography{lit}

\end{document}